# Stability and superconductivity of K-Y-H hydrides under high pressure

Ling-Yan Chen，Wen-Cai Lu, Hai-Liang Chen, Bing-Yu Li, Kai-Ping Yang

*College of Physics and State Key Laboratory of Bilological Polysaccharide Fiber Forming and Ecological Textile, Qingdao University, Qingdao, Shandong 266071, PR China*

*Institute of Theoretical Chemistry, Jilin University, Jilin 130021, P. R. China*

**Abstract**

It has been a hot issue to search for hydrides at high pressure with high superconducting transition temperature (Tc). In this work, ternary hydrogen-rich compounds $KYH_n$ (n = 1-16) were studied and calculated using the density function theory (DFT) method and he results showed that the $KYH_8$ demonstrates good stability and superconductivity. And keep thermodynamically stable in the pressure range of 50-300 GPa. The research found that the Tc of $KYH_8$-C2/m was predicted to be 111.3K, 122.0K, and 142.3K at 200, 250 and 300 GPa, respectively. Our current study provides a possibility for searching new high-Tc superconductors in ternary hydrides.

## 1. Introduction

The search for high-temperature superconductors is still an exciting subject, as its wide application has aroused great interest among scientists. Based on the BCS electron-phonon superconducting theory, high frequency phonons, strong electron-phonon interactions, and large electron state density (DOS) near the Fermi level contribute to the high superconducting transition temperature Tc. Hydrogen is the lightest element in nature and promotes high phonon frequency and electron-phonon coupling. Theoretical predictions show that when solid hydrogen turns to metal at high pressure, they are very likely to become high temperature superconductors. The concept of metal superconductivity has been extended to hydride-rich compounds. In 2004, Ashcroft [5] predicted that "chemical precompression" could be caused by heavier elements, metallize hydrogen-rich materials at much lower pressures than pure hydrogen. Theoretical and experimental results show that many hydrogen-rich compounds, such as $H_3S$[10], $H_2S$, $LaH_{10}$[18], $YH_{10}$, $YH_6$,[27] $CaH_6$ ,[29]

$MgH_6$, exhibit high Tc under high pressure. Theoretical research predicts that when $H_3S$ is in 200GPa, Tc can reach 191-204K[10] with Im-3m phase. And the prediction was confirmed in subsequent experiments. For the alkali metal compounds, $CaH_6$ with a cage structure at a high pressure had a predict Tc of 235 K at 150 GPa, $MgH_6$ critical temperature of ~260 K above 300 GPa and $KH_6$[8] superconducting transition temperature in the range of 58.6 K to ~69.8 K at 166 GPa. In addition, the theoretically predicted that $YH_3$ is most stable in thermodynamics, has a Tc of 40 K at 17.7 GPa[3], $YH_6$ and $YH_{10}$, the H atoms form a sodalite-like cage and $H_{32}$ cage[13], respectively, have critical temperature of 251-264 K at 120GPa and 305-326 K at 250 GPa. $LaH_{10}$ had Tc of 274-286 K at 210 GPa.

Recently, ternary hydrides have been studied. The superconducting properties of several kinds of ternary hydrides have been reported. At 100GPa, $BaReH_9$[11] has been reported to be a superconductor with a Tc of ~7 K. In 2017, $MgGeH_6$[12] was reported to be a potential high temperature superconductor with a Tc of ~67 K at 200 GPa. In the near future, theoretical research reveals that $CaYH_{12}$ becomes stable with a cubic Fd3m structure above 170 GPa, with Tc of ~258 K at 200 GPa[28]. Then, calculation shows that $KScH_{12}$ with a hydrogen cage has critical temperature of 122 K at 300 GPa, and $GaAsH_6$ with a Tc of~ 98 K at 180 GPa.

In this paper, we studied the ternary hydrides of $KYH_n$ at a pressure range of 100 to 300 GPa. We used genetic algorithm (GA) to search for the structure of KYHn and then studied their thermodynamic stabilities, electronic properties and superconductivity. The most stable structure of $KYH_8$ is cubic c2/m, it was predicted Tc to be 0.3K at 50GPa, 111K at 200GPa, and 141K at 300GPa.

2. computation details

The low-enthalpy structures of KYHn (n=1-16) at high pressure were searched by genetic algorithm (GA) combined with structural optimizations using density functional theory (DFT). We used the Perdew-Burke-Ernzerhof (PBE) Generalized gradient approximation (GGA) [4] functional and Norm-conserving pseudopotentials, implemented in the CASTEP code are employed. The GA search of low-enthalpy structures of KYHn were performed at 100 GPa. In the GA search, a coarse optimization method was used with an energy cutoff of 330 eV and a k-mesh of 5 ×6 ×7. Then the structures from the GA search were re-optimized with high precision at the pressure range 50 - 300 GPa, with energy cutoff of 1000 eV and a k-point mesh of 2π × 0.03 Å-1 to determine low-enthalpy structures. The Quantum-ESPRESSO package [6] was used to calculate phonon, electron-phonon coupling (EPC) and Tc, using the PBE functional, Norm-conserving pseudopotential and an energy cutoff of 70 Ry.

## 3. Results and discussion

### 3.1 Structural stability of K-Y-H

The stable structures of $KYH_n$ were studied at the pressure range of 100-300 GPa. Unlike binary compounds, the synthesis and decomposition pathways of ternary compounds are more complicated.

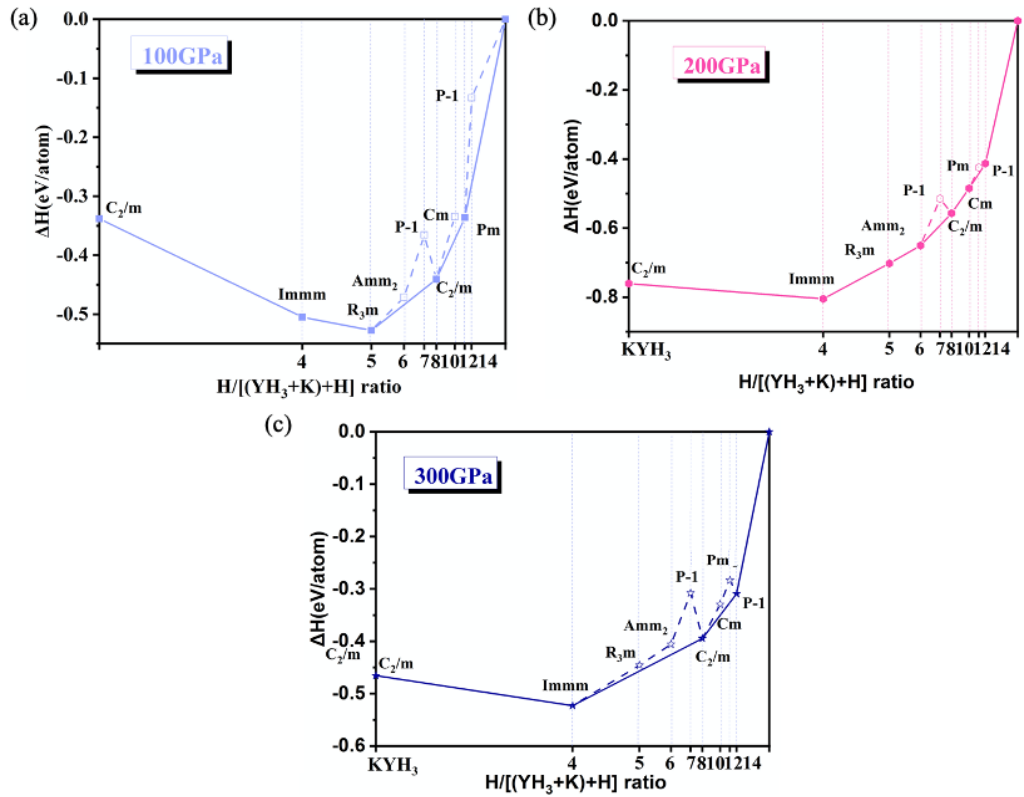

Figure 1 Calculated formation enthalpies (ΔH) of various K-Y-H stoichiometries at 100, 200 and 300 GPa with respect to $(K+YH_3)+ 1/2$ solid $H_2$. The solid lines represent convex hull and solid symbols on convex hull indicate stable stoichiometries. Hollow symbols connect by dashed lines represent unstable stoichiometries

The calculated formation enthalpies of the KYHn hydrides at 100, 200 and 300 GPa with respect to (K+YH$_3$) + 1/2 solid H$_2$. It can be seen that KYH$_5$ is most stable at 100 GPa, and KYH$_4$ is most stable at 200 − 300 GPa .The KYH3,4,5,8.12 at 100 GPa, KYH3,4,5,6,8,10 and 12 at 200 GPa, KYH3,4,8 and 14 at 300 GPa locate on the convex hull meaning they may be thermodynamically stable (Figure 1). We extend the structure to 2 f.u. Among them, the crystal structures of KYH8-C2/m and KYH$_{14}$-P-1 at 300GPa were shown in Figure 2. In the figure2, KYH$_8$-C2/m and KYH$_{14}$-P-1 in the directions of 001,010,100 are shown, in which the pink atom is H, the yellow atom is K, and the blue atom is Y. For KYH$_8$, we can observe the presence of the H$_3$ unit in the structure and the heavy atom Y combined with two hydrogen to form a straight chain structure, in which the H$_3$ unit is likely to be the source of high temperature superconductivity. There are two types of H-H bonds in the crystal, with distances of 0.921Å (H1-H2), 0.819Å (H$_3$-H$_5$), and the bond length between H$_7$ and Y is 1.77338Å (Figure 2). Their detaileds were given in Table S2 in the supplementary material.

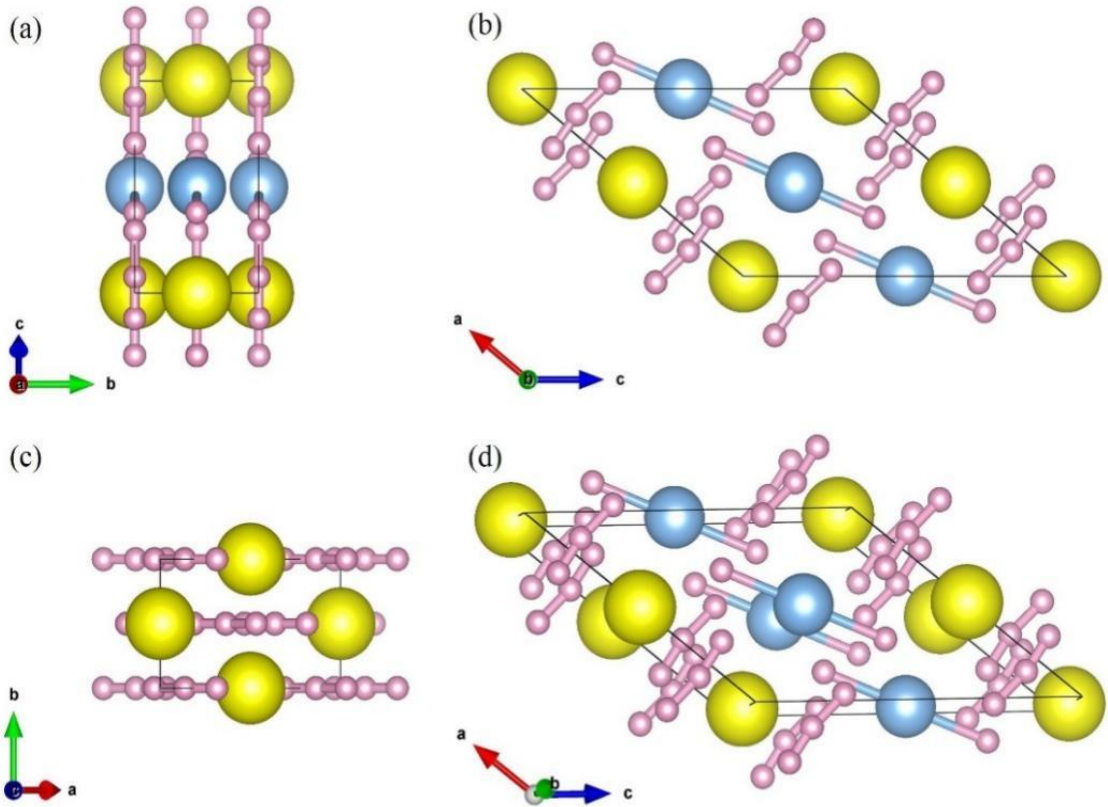

We caculation the electronic localization function of C2/m KYH$_8$ at 300 GPa to explore the bonding information. The ELF has maps values in the range from 0 to 1, with 1 corresponding to excellent localization of valence electrons, standing for a strong covalent bond. The ELF values for the $H_1$-$H_3$ and $H_3$-$H_5$ bonds are nearly 0.9, meaning they are strong covalent bonds. The ELF value of the Y-$H_7$ bond in the octagon is approximately 0.5, which is the corresponding value for the homogeneous electron gas (Fig. 3(a)). Fig. 4 shows the calculated relative enthalpy curves for the predicted structures of $KYH_8$ as a function of pressure. The enthalpies of the decomposition to binary hydrides were also taken into account. On the basis of corresponding binary compounds, the decomposition enthalpies into K+Y+$4H_2$, K+YH3+$5/2H_2$, K+$YH_6$+$H_2$, and $KH_6$+Y+$H_2$ are considered in order to investigate the stability of KYH8. For KYH8, four energetically competing phases C2/m, Pmm2, Amm2, and Cm were found by our structural predictions. The crystal structures and lattice parameters are shown in Fig. S3. Without considering zero-point energy (ZPE), we found that the c2/m is the most stable phase below 90GPa. Above 93GPa, the phase of Amm2 becomes thermodynamically stable above 90 GPa. The results can show $KYH_8$ is stable with respect to K+$YH_3$+$5/2H_2$ in the pressure range, and can still table with respect to $KH_6$+Y+$H_2$ below 280 GPa.

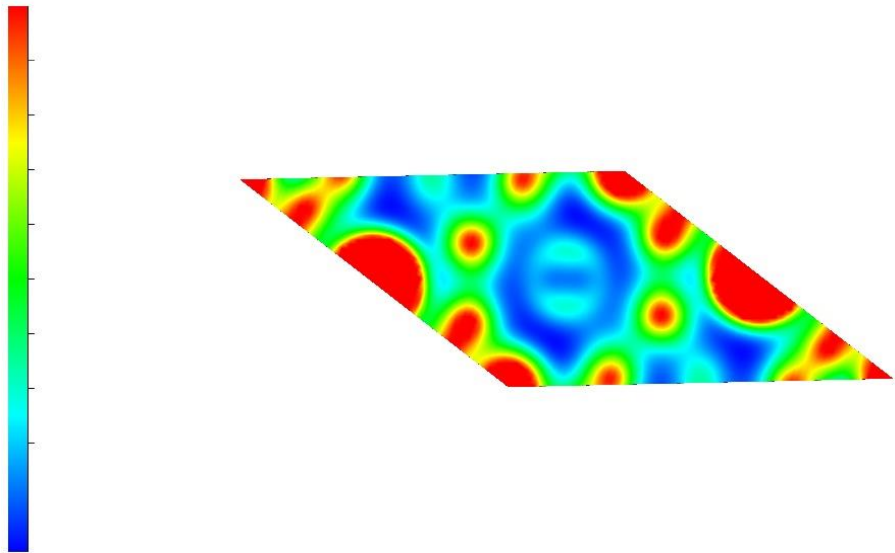

Figure 3 The electronic localization function of C2/m KYH8

3.2 Electronic properties, dynamic Properties and superconductivity of KYH

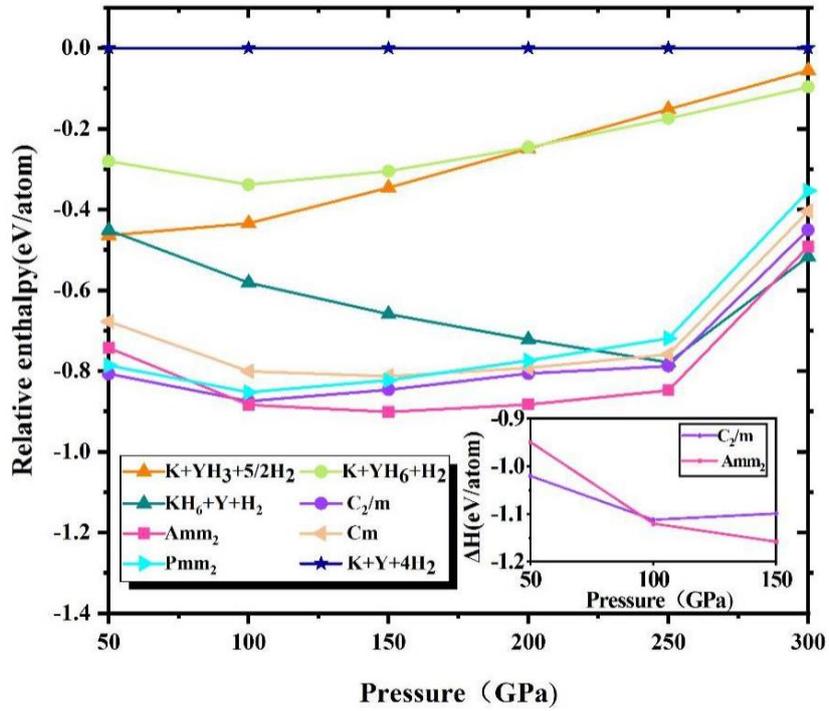

Figure 4 The calculated enthalpies of various structures for (a) $KYH_8$ as a function of pressure with respect to the $K+Y+4H_2$. The decomposition enthalpies for $KYH_8$ to $K+YH_3+5/2H_2$, $K+YH_6+H_2$, $KH_6+Y+H_2$ structure including zero-point corrections as a function of pressure.

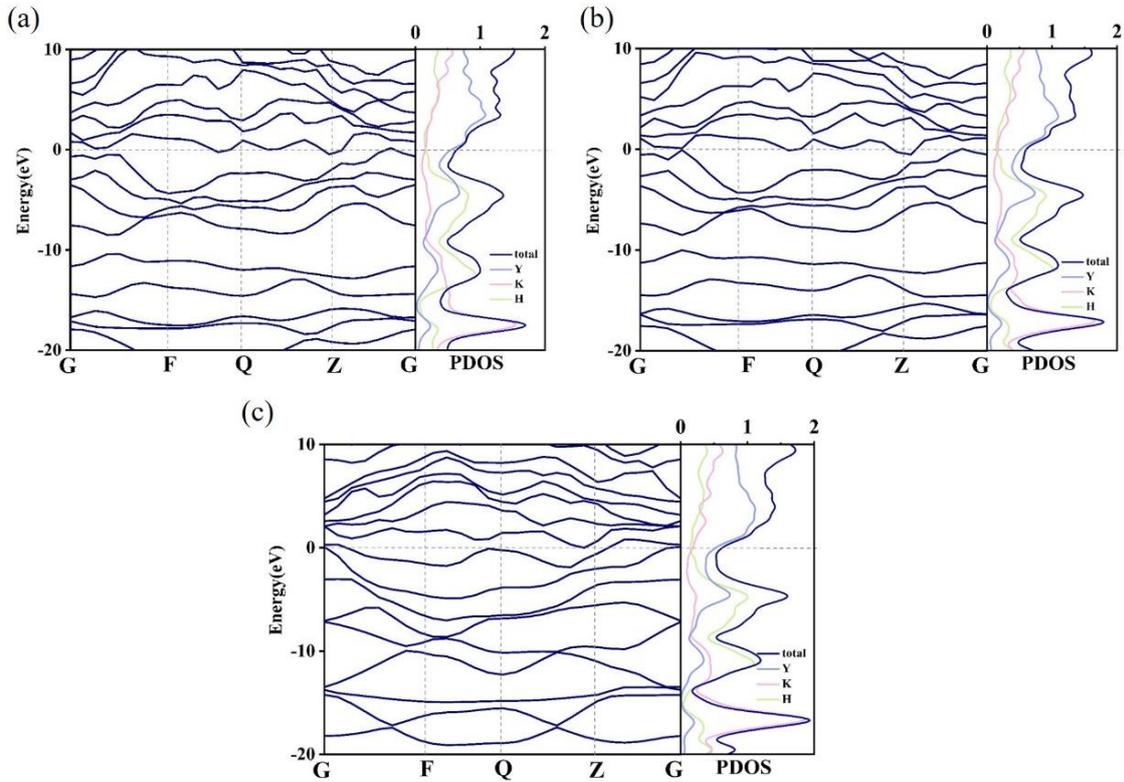

Figure 5 Calculated electronic band structures and partial density of states (PDOSs) of (a) $KYH_8$-c2/m at 200 GPa, (b) $KYH_8$-c2/m at 250 GPa, (c) $KYH_8$-c2/m at 300 GPa,

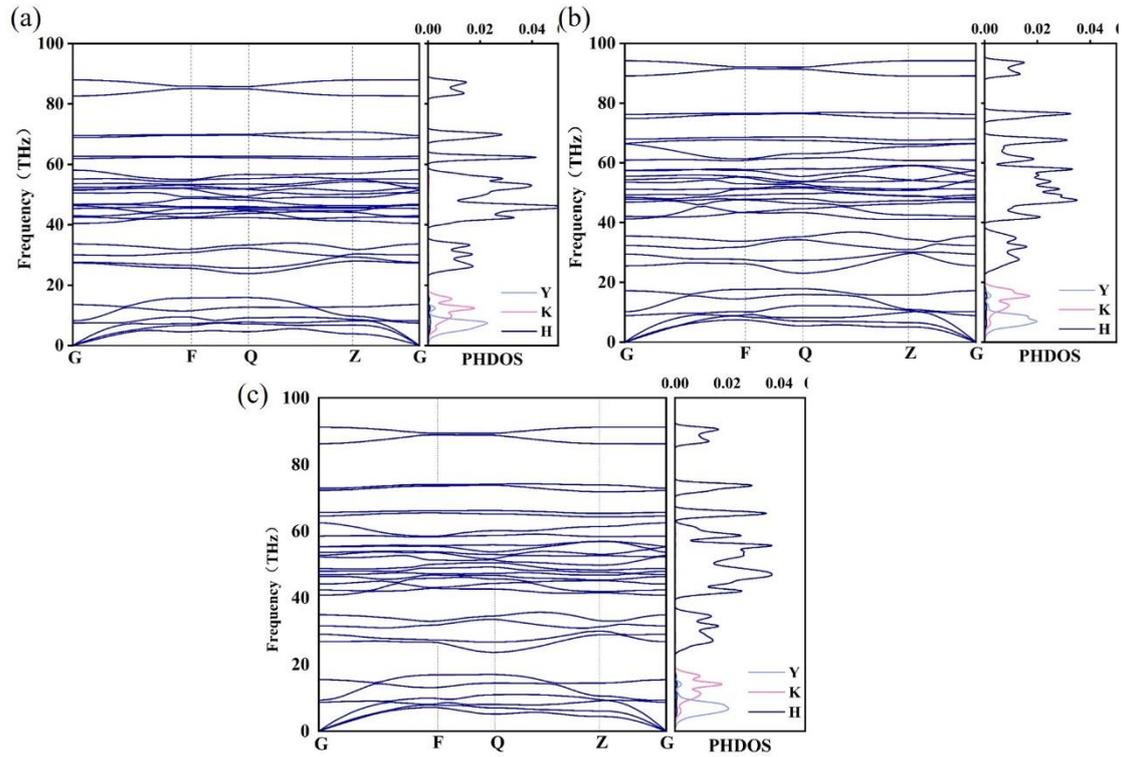

Figure 4 Calculated phonon dispersion curves, phonon density of states (PHDOSs) of $KYH_8$-C2/m (a) KYH8-c2/m at 200 GPa, (b) KYH8-c2/m at 250 GPa, (c) KYH8-c2/m at 300 GPa,

For $KYH_8$, we further research their electronic dynamic and their superconducting properties. In order to confirm whether they are metals under the pressure, we analyzed the electronic band structures and projected density of states (PDOS) for the predicted structures. As shown in Figure 4 (a), (b)and (c), There are many bands crossing the Fermi level, the band gaps of KYH8-c2/m at 200, 250 and 300GPa are zero, indicating they are metallic. the overlap of the valence band and the conduction band in the Fermi level reflects the metallic properties of the structure. The energy band on the Fermi level can effectively affect the critical temperature. There are some flat bands around the Fermi level, which may contribute to enhance electron-phonon interactions. To explore the electronic, dynamic and mechanical properties of KYH8, we calculated the phonon dispersion curves, phonon density of states (PHDOS) of KYH8-C2/m and. The results indicate that there phonon dispersion curves has no virtual frequency (Figure5), which can help us determine their structures are dynamic stable. As shown in the phonon DOS, the vibrations of the heavier Y and K atoms are contributed to the low-frequency phonon branches, while vibrations of H atoms are associated with the high frequency regions, respectively. Based on the stability of the structure at various pressures, we calculated Eliashberg spectral function $\alpha^2F(\omega)$ and electron phonon coupling (EPC) parameters $\lambda(\omega)$ of $KYH_8$-c2/m to further explore the superconducting properties. There are two parameters that can be used to estimate Tc ( $T_c = \frac{\omega_{log}}{1.2}\exp[-\frac{1.04(1+\lambda)}{\lambda-\mu^*(1+0.62\lambda)}]$), calculating electron phonon coupling (EPC) parameter $\lambda$ and the logarithmic average phonon frequency $\omega$log, and Eliashberg phonon spectral function $\alpha^2F(\omega)$ and $\lambda$ are shown in Fig. 6

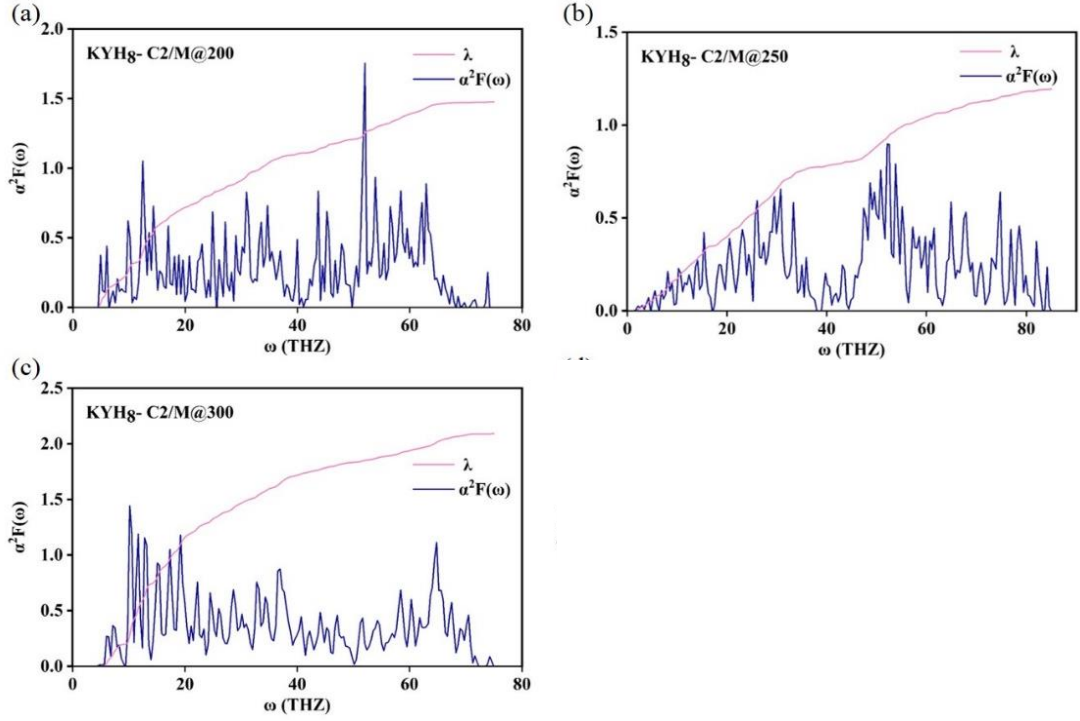

Figure 6 Eliashberg phonon spectral function $\alpha^2F(\omega)$, and electron–phonon coupling parameter $\lambda$ of (a) $KYH_8$-c2/m at 200 GPa, (b) $KYH_8$-c2/m at 250 GPa, (c) KYH8-c2/m at 300 GPa,

Table 1 Calculated EPC parameter $\lambda$, $\omega$log, DOS at the Fermi level N($\varepsilon$f), and Tc at different pressures.

|  | Pressure (GPa) | $\lambda$ | $\omega_{log}$(K) | N ($E_f$) | Tc(K) $\mu^*$=0.1 | Tc(K) $\mu^*$=0.13 |
|---|---|---|---|---|---|---|
| C2/m-$KYH_8$ | 200 | 1.478 | 989.897 | 2.76 | 111.3 | 101.4 |
|  | 250 | 1.214 | 1342.505 | 4.47 | 122.0 | 108.3 |
|  | 300 | 2.09 | 960.470 | 0.65 | 142.3 | 133.2 |

And then, we calculated the electron-phonon coupling (EPC) parameter $\lambda$, the phonon frequency logarithmic average ($\omega_{log}$) and DOS at the Fermi level N$_{(\varepsilon f)}$ by QE software package. We chose the Coulomb pseudopotential $\mu*$ of 0.1 and 0.3, the

estimated Tc values for C2/m-KYH$_8$ are 111.3K, 101.4K, 142.3K and 122.0K, 108.3K, 133.2K at 200, 250 and 300GPa, respectively (Table1). It is obvious that the transition temperature increases with pressure.

## 4. Conclusions

In summary, the crystal structures and stability of the KYHn under high pressure were searched by GA and combined with the structural optimizations with DFT. And we systematically studied the structures and stabilities of KYHn at the pressure range 50-300 GPa. It is found that KYH8 c2/m have strong electron phonon coupling and a high Tc (142.3K and 152.3K at 300GPa). The H3 unit is found in the structure of KYH$_8$ c2/m likely to lead to a higher Tc.

**Declaration of competing interest**

The authors declare that they have no known competing financial interests or personal relationships that could have appeared to influence the work reported in this paper.

**Acknowledgments**

This work was supported by the National Natural Science Foundation of China (Grant No. 21773132).